\magnification=1200
\parskip = 0pt
\baselineskip = 15 true pT
\hsize 16.5 true cm
\vsize 22.6 true cm
\raggedbottom
\pageno = 1

\footline={\hfil}
\headline={\ifnum\pageno=1 \hfil \else\hfil\folio\hfil\fi}

\def\delv{$\Delta V^{HB}_{TO}$}

\def\etal {{\it et al.}\ }

\def\eg {{\it e.g.},\ }
\def\cf {{\it cf.}\ }
\def\ie {{\it i.e.},\ }

\def\Teff {T_{e\!f\!\!f}}

\def\dm0 {(m-M)_0}

\def\BV0 {(B-V)_0}
\def\ngc#1{{NGC$\,$\hbox{#1}}}
\def\M#1{{M$\,$\hbox{#1}}}
\def\lta{\mathrel{\hbox{\raise 0.6 ex \hbox{$<$}\kern
                   -1.8 ex\lower .5 ex\hbox{$\sim$}}}}
\def\gta{\mathrel{\hbox{\raise 0.6 ex \hbox{$>$}\kern
                   -1.7 ex\lower .5 ex\hbox{$\sim$}}}}

\line{~~~\hfill}
\vskip 0.5 true in
\centerline{THE RELATIVE AGES OF GALACTIC GLOBULAR CLUSTERS}
\vskip 0.3 true in
\centerline{Peter B. Stetson}
\smallskip
\centerline{Dominion Astrophysical Observatory, Herzberg Institute of
Astrophysics,}
\centerline{National Research Council of Canada,}
\centerline{5071 West Saanich Road, Victoria, B.C., Canada\ \ V8X~4M6}
\bigskip
\centerline{Don A. VandenBerg\footnote{$^1$}{Killam Research Fellow}}
\smallskip
\centerline{Dept. of Physics \& Astronomy, University of Victoria,}
\centerline{P.\ O.\ Box 3055, MS 7700, Victoria, B.C., Canada\ \ V8W~3P6}
\bigskip
\centerline{and}
\bigskip
\centerline{Michael Bolte}
\smallskip
\centerline{UCO/Lick Observatory, University of California,}
\centerline{Santa Cruz, California, U.S.A.\ \ 95064}
\vfill
\vfill
\noindent Shortened Title: RELATIVE GLOBULAR CLUSTER AGES
\vfill
\line{Send proofs to:~~Peter B. Stetson\hfil}
\line{~~~~~~~~~~~~~~~~~~~~~ Dominion Astrophysical Observatory\hfil}
\line{~~~~~~~~~~~~~~~~~~~~~ 5071 West Saanich Road\hfil}
\line{~~~~~~~~~~~~~~~~~~~~~ Victoria, B.C., CANADA\ \ \ V8X~4M6\hfil}
\line{~~~~~~~~~~~~~~~~~~~~~ Phone: (604)~363-0029\qquad Fax: (604)~363-0045\hfil}
\line{~~~~~~~~~~~~~~~~~~~~~ E-mail: stetson@dao.nrc.ca\hfil}
\vfill
\line{~~~~~~~~~~~~~~~~~~~~~ Co-authors' e-mail: davb@uvvm.uvic.ca, bolte@lick.ucolick.org\hfil}
\vfill\eject
\baselineskip = 25 true pt

\centerline{ABSTRACT}

\noindent We discuss the present state of knowledge and thought concerning
the spread in age found among Galactic globular clusters, with some
discussion of the implications for what happened during the earliest
stages of the formation of the Milky Way Galaxy.  Differential
observational techniques to derive the relative age differences among
clusters of similar metallicity are discussed in detail.  We conclude that
as of the current date (mid-April 1996) the state of the field is still
somewhat muddled.  However, we believe that there is now a substantial
body of evidence --- including a particularly revealing intercomparison of
the color-magnitude diagrams of \ngc{1851}, \ngc{288}, and \ngc{362}
presented here --- indicating that age is {\it not\/} the dominant second
parameter determining the shape of globular clusters' horizontal
branches.  If our assertion is correct, then apart from a handful of
anomalous clusters that may well have been captured from a satellite dwarf
galaxy, there is no strong evidence either for a significant spread in age
among clusters of a given metal abundance or for a systematic variation of
mean age with Galactocentric distance.  On the question of whether there
is a significant age difference between metal-poor and metal-rich
clusters, we feel compelled to fall back on the Scottish verdict:
``Unproven.''  Data now being collected by numerous groups in various
subdisciplines may resolve the remaining controversy within a few years.

\bigskip

\centerline{1.\ INTRODUCTION}

Few numbers in science are more avidly sought than the ages of the
Galactic globular clusters (GCs).  These spherical, compact,
gravitationally bound systems of typically $10^5$--$10^6$ stars that orbit
the center of the Milky Way (and most, if not all, other large galaxies),
are believed to be among the first recognizeable stellar systems that
formed in the Universe (see Peebles and Dicke 1968).  As they are the most
ancient objects known for which reliable ages can be determined, and as
the Universe cannot be younger than the oldest object it contains, the GCs
provide one of the most important of the few constraints that we have on
cosmological models.  In tandem with this study, VandenBerg, Bolte, and
Stetson (1996) review our present understanding of {\it absolute\/}
cluster ages and briefly assess their implications for standard Friedmann
cosmologies.

The calibration of absolute ages is still subject to observational and
theoretical uncertainties at the $\approx 20$\% level, and represents a
major limitation on our ability to test cosmological models.
Nevertheless, it is quite possible to determine {\it relative\/} GC ages
with sufficient precision to address some outstanding questions concerning
the formation of our own Milky Way Galaxy.  In particular, we now have the
means to estimate --- to within a fraction of a Gyr --- the dispersion in
age of globular clusters having very similar chemical compositions, as
well as the variation of age with Galactocentric distance encompassed by
(once again) those systems of comparable metal abundance.  Even though the
derivation of a reliable age--metallicity relation for the GCs remains
problematic, due largely to distance-scale uncertainties, we may soon be
able to ascertain whether the protoGalaxy underwent the type of rapid,
monolithic collapse envisioned by Eggen, Lynden-Bell, and Sandage (1962),
or whether the Galaxy formed much more gradually through the accretion of
independent sub-systems over a prolonged period of time (cf. Searle and
Zinn 1978).  It is the purpose of this investigation to review the many
advances that have been made in recent years towards measuring relative
ages of GCs with a precision sufficiently high ($\lta 1$Gyr) to resolve
the first epoch of star formation in the Galaxy. Of course, there are many
other paths of investigation that are leading toward a complete
understanding of our Galaxy's childhood and adolescence based on
studies of halo and thick disk field stars; investigations of the
Sagittarius galaxy, Carina, and other nearby dwarfs; and {\it in situ\/}
observations of galaxies ``now'' forming at large redshift.  While we will
not discuss these approaches in this review, we do not in any way wish to
minimize their importance.

\bigskip

\centerline{2. BACKGROUND: CONCEPTS OF HOW THE GALAXY FORMED}
\nobreak
The general notion that the Galaxy reached its present form after a
self-gravity-driven collapse from a more dispersed state dates back at
least to Kant (1755).  This classical view of the formation of the Galaxy
has been given a modern codification and quantification by Eggen,
Lynden-Bell, and Sandage (1962, hereafter ELS), who proposed that the
original protoGalaxy consisted of a single large gas cloud with a radius
(in present-day coordinates) of order 100~kpc.  Unable to support itself
by gas pressure once its mass had stopped the local Hubble expansion, this
cloud immediately underwent a free-fall collapse.  The globular clusters
and field halo stars are presumed to have formed in condensations embedded
within the general intraGalactic medium during this rapid free-fall phase,
whose duration could not have exceeded a few times $10^8$ years.  Finally,
when the gas achieved a density which allowed pressure support, and
dissipation therefore began to become important, it was able to radiate
away its kinetic energy;  residual angular momentum then forced it to
settle into a rotating disk, where all subsequent star and cluster
formation has taken place.  In contrast to the gas, the dissipationless
halo field stars and GCs have retained the kinetic energies, angular
momenta, and general spatial distribution that they possessed at the
instant the gaseous collapse ceased.

The principal alternative to this model was presented by Searle and Zinn
(1978, hereafter SZ).  Arguing from the postulate that differences in age
were responsible for the range in horizontal-branch types at fixed metal
abundance --- the so-called ``second parameter'' effect (van~den~Bergh
1965, 1967; Faulkner 1966; Sandage and Wildey 1967) --- they inferred that
the formation of the Galactic halo could {\it not\/} have been as rapid as
a free-fall collapse.  According to early horizontal-branch models (\eg
Faulkner 1966; Rood and Iben 1968; Simoda and Iben 1970; Castellani and
Tornamb\`e 1977), the observed range in horizontal-branch morphological
types would require that the epoch of globular-cluster formation must have
been $\gg$1~Gyr.  SZ achieved the required prolonging of the cluster
formation time in the outer halo by abandoning the notion of a continuous
medium in free-fall radial collapse.  Instead, they adduced the
theoretical models of Binney (1976), which assumed that the protoGalactic
material consisted of distinct ``flows'' of gaseous material which would
not be directed purely in the radial direction, but which would on
occasion collide with each other, dissipating their kinetic energy by
shock-induced radiation and producing transient sheets of cool, compressed
gas suitable for the formation of a generation of stars and clusters.  If
the filling factor of these gas flows was sufficiently small, the time
between collisions could be long enough to spread out the dissipation of
the original kinetic energy over the time span required to produce the
inferred age spread in the outer halo.

The relationship between these transient sheets forming as a result of
collisions between currents of protoGalactic gas, and the ``protogalactic
fragments'' discussed earlier by Searle (1975), is not entirely clear.
Searle's earlier paper expands upon previous discussions by von~Weizsacker
(1955) and Oort (1958), who may have been the first to suggest that the
collapse of the Galaxy was not the simple radial, gravity-driven collapse
envisioned by Kant and by ELS, but rather that turbulent motions within
the protoGalactic material moderated the rate of contraction through the
gradual dissipation of kinetic energy.  Searle originally adopted the
speculations of von~Weizsacker and Oort to explain why the range of metal
abundance among stars within a given GC is immeasureably small compared
with the total chemical enrichment in the typical cluster.  This
circumstance would be extremely hard to understand if the material which
became a given globular cluster separated itself out from the general
protoGalactic medium early in the free-fall phase, and then enriched
itself in heavy elements through successive supernova events. One would
then expect to see a range of metal abundances from zero up to some
maximum, representing stars that formed in pockets of variously enriched
gas during the intervals between the supernova explosions.  Instead,
Searle suggested that the protohalo consisted of ``initially gaseous
fragments, each small compared with the Galaxy but large compared with
globular clusters.\ $\ldots$ Following Kaufman's (1975) discussion of the
efficacy of mixing by supernova explosions, it seems reasonable to assume
complete mixing within a fragment and no mixing between them.'' Thus,
within a given fragment, initial generations of stars would produce
chemical enrichment and complete remixing of those secondary elements
throughout the mass of the fragment.  At various times, star clusters
would form, each inheriting the unique metallicity belonging to the parent
fragment at that instant.  Each fragment would eventually encounter the
main body of the Galaxy; its gaseous content would be stripped and added
to the Galactic disk, while its stars and clusters would return to the
halo.  Subsequently, Searle and Zinn proposed the very similar (though not
identical) scenario already discussed to enable them to adopt a very
different postulate: namely, that the second parameter is age.  The similarity
of the two situations envisioned by Searle and by SZ has led some authors
--- including, on occasion, us --- to confuse the two, to speak of
``Searle-Zinn fragments'' and to suggest that the present-day dwarf
spheroidal galaxies may be the aged remains of a few of these primordial
lumps of protoGalactic material that have managed to survive to the
present day.  So does the model really envision many independent long-lived
condensations zipping through otherwise empty space like bees buzzing
around a hive, all the while undergoing continuous internal evolution
until, one by one, they collide and merge their gaseous bodies to form the
kinematically ordered Galactic disk, while their collisonless stellar and
cluster progeny continue to roam the original spherical volume once
occupied by the whole protoGalaxy?  Or were the stars and clusters formed
in transient sheets of cool material behind the shock fronts of colliding
gas streams, sheets which formed but one generation of stars and clusters
before they were ionized by OB stars and torn apart by supernovae and
returned to the turbulent, gaseous, spheroidal protoGalaxy --- a process
that has nothing whatever to do with modern dwarf galaxies?  Even Oort
spoke alternatively of ``irregular streamings'' and of ``gas clouds'' which
undergo collisions; it is not clear that Searle and Zinn specify which
version they prefer, either.

More recently, the detailed theoretical modelling of horizontal branch
morphologies by Lee, Demarque, and Zinn (1988, 1994; also Lee 1991, 1992,
1993) have suggested that the time span required for the formation of halo
objects is $\gta$5~Gyr; in particular, Lee, Demarque, and Zinn (1988)
inferred an age difference of 6~Gyr for the well-known second-parameter
pair of globular clusters, \ngc{288}\ and \ngc{362}, if the absolute age
of the older of the two (\ngc{288}) is 18.5~Gyr --- a
difference-over-average of $\sim$40\%.  In their more recent discussion,
Lee, Demarque, and Zinn (1994) revised the estimated difference to
3--4~Gyr for an adopted age of 14.9~Gyr for \ngc{288} --- a difference of
$\sim$30\%.

Sandage (1990) has argued for what is essentially a synthesis of the two
models.  He contends that critics of ELS have been overly literal in their
interpretation of its highly simplified picture, as naturally there would
necessarily have been some variation of local densities within the
protoGalactic gas cloud, and denser lumps would have collapsed in
upon themselves on shorter timescales even as the Galaxy overall collapsed
in upon {\it it\/}self on a longer timescale.  Still, there remains a
dichotomy of opinion, with some agreeing with Sandage that the collapse
was ``coherent'' (albeit noisy), ``rapid'' (in the sense that the radial
infall greatly exceeded turbulent motions, as required by the highly
eccentric orbits of the halo stars), and dominated by gravity; while
others agree with SZ that, at least in the outer halo, the collapse was
``chaotic'' and ``slow'' (required by the age interpretation of the
second-parameter problem), and that gravity initially dominated over
kinetic energy by only a tiny margin.  The most extreme version of the
latter picture is probably that of Zinn (1993), who argued that the
``young'' globular clusters in the outer halo had their origin in just a few
satellite galaxies (resembling, perhaps, the Magellanic Clouds) which have
only quite recently been disrupted and accreted by the Galaxy.

It must be recalled that, up to 1987, the age explanation of the
second-parameter phenomenon was still largely a hypothesis, relying upon
assumptions concerning the universality of chemical abundance ratios, the
internal angular momentum of stars, mass-loss mechanisms, and the like.
None of the so-called ``second-parameter'' clusters had had its
chronological age accurately measured by the more fundamental techniques
based on the luminosity and temperature of the main-sequence turnoff.
But, beginning in 1987, such measurements began to appear in rapid order.

\bigskip

\centerline{3.\ MODERN MEASUREMENTS OF RELATIVE CLUSTER AGES}
\nobreak
The true measure of a star cluster's age is the intrinsic luminosity of
its main-sequence turnoff.  A star can be thought of as a gravitationally
confined reservoir of nuclear fuel.  During nearly all of its lifetime,
the luminosity of a star is set by the requirement that it support itself
against gravitational contraction, and a new, young star quickly adjusts
its internal density and temperature structure until the fusion of
hydrogen to helium in the stellar core provides precisely the required
rate of energy production.  Over the approximately 90\% of a star's total
lifetime that is spent in this so-called ``main-sequence'' phase of its
evolution, it grows gradually more luminous and slightly hotter in
response to the changing chemical structure of its interior.  There comes
a time in the life of a star, however, when it has exhausted the available
hydrogen fuel at its center.  Then the inert helium core, containing some 10\%
of the star's mass, shrinks to a volume of order the dimensions of the
planet Earth while the outer layers of the star rapidly expand in radius,
causing the surface temperature of the star to fall precipitously as
hydrogen continues to be consumed in an energy-generating shell around the
helium core.  The rate of fuel consumption during the main-sequence phase
of evolution is known from the star's luminosity; the total amount of fuel
available in the hydrogen-burning core is easily estimated from even the
most primitive stellar interior models; and the ratio of these two numbers
yields the amount of time a star of a given mass can remain on the main
sequence.  More massive stars consume their fuel far more rapidly than
less massive stars, and die proportionately sooner.  A star cluster
represents a coeval assemblage of stars spanning a range of mass: by
simply observing a cluster and determining the mass or luminosity of stars
that are just switching over from slow hotward evolution to rapid coolward
evolution, one has a direct measure of the age of the cluster.

Although the commonly held wisdom through the mid-1970's was that globular
clusters were coeval, this interpretation was based on photoelectrically
calibrated photographic photometry.  The difficulties of photometering
main-sequence-turnoff and fainter stars on photographic plastes were
formidable. In retrospect it is clear that even age dfferences as large as
3 -- 4 Gyr could have been lost in the random and systematic photometry
errors.  However, when CCDs came into common use in the mid-1980s and the
software for deconvolving overlapping stellar images in the invariably
crowded GC fields became widely available, the stage was set for
establishing relative GC ages with uncertainties $<1$~Gyr.

The first clear measurement of an anomalously young age for a Galactic
globular cluster emerged from two studies of the outer-halo,
second-parameter cluster Palomar~12, which Ortolani (1987) reported to be
younger by some substantial (but unspecified) amount than the other
Galactic GCs; at the same conference, Stetson and Smith (1987)
independently reported an anomalously young age for Pal~12, and estimated
by a differential comparison with 47~Tucanae and \M5\ that it was some
30\% younger than those two nominally normal clusters.  Subsequent more
exact analyses (Gratton and Ortolani 1988; Stetson \etal\ 1988, 1989)
supported this provisional estimate.  Since then, the clusters Ruprecht
106, Terzan~7, Arp~2, and IC 4499 (Buonanno \etal\ 1990, 1993, 1994, 1995;
Kubiak 1991; Da~Costa, Armandroff, and Norris 1992; Ferraro \etal\ 1995)
have been added to the roster of clusters unequivocally younger than the
majority.  However, as discussed by Lin and Richer (1992) and Fusi~Pecci
\etal\ (1995), it is possible that these clusters have been stripped from
the Magellanic Clouds by the tidal force of the Galaxy, or from another
satellite galaxy that has since been disrupted.  Indeed, the recent
discovery of the dwarf spheroidal galaxy in Sagittarius (Ibata, Gilmore,
and Irwin 1994), and the possibility that the globular clusters
\ngc{6715}\ = \M{54}, Arp~2, Terzan~8, and conceivably Terzan~7, may be
associated with it, could be a contemporary example of this process.
Therefore it is unclear whether these clusters trace the dominant
formation mode of the original halo, or just a small minority component of
it.

To understand the principal formation mode of the Galactic halo, it will
be necessary to measure precise relative ages for a majority of the
globular clusters.  Work to date tends to suggest that the age dispersion
among the nearest clusters is small, but perhaps perceptible.  Bolte
(1989) obtained precise relative photometry for the main-sequence turnoffs
of the well-known nearby second-parameter pair \ngc{288}\ and \ngc{362},
from which he concluded that the former was older than the latter by some
3~Gyr.  Green and Norris (1990) confirmed this result shortly thereafter
(however, see \S 4 below).  Note that this age difference is appreciably
smaller than the one originally suggested by Lee \etal\ (1988): they had
inferred an age of 18.5 Gyr for \ngc{288}, versus 11.2 Gyr for \ngc{362}.
Lee (1991) subsequently pointed out that his previous estimate of the age
difference had been made on the basis of a Solar ratio of oxygen to iron
in the globular-cluster stars, and that allowing for various effects
related to an enhancement of oxygen would reduce the absolute size of the
age difference.  Their most recent calibration of theoretical
horizontal-branch morphologies (Lee \etal\ 1994) yields a relative age
difference of 3--4~Gyr, if the absolute age of \ngc{288} is 14.9 Gyr.
This would appear to be a firm lower limit to the difference in age
between these two clusters if age is the sole cause of the disparity in
their respective HB morphologies.  Catelan and de~Freitas~Pacheco (1993,
1994) have found from their HB simulations that an age difference as small
as $\approx 2$--3 Gyr would require that either both clusters are $\lta
10$ Gyr old or some parameter {\it in addition\/} to age is different
between them.

VandenBerg, Bolte, and Stetson (1990; hereafter VBS) compared the
color-magnitude diagrams (``CMDs'') of another nearby second-parameter
pair, \ngc{5272}\ = \M{3}\ and \ngc{6205}\ = \M{13}, and estimated an age
difference of some 1--2~Gyr (7--15\%) as compared to the Lee \etal\ (1994)
prediction of $\approx 3$ Gyr ($>$20\%) from their synthetic HB
calculations.  Once again, it would seem to be impossible for modern HB
models to accommodate a smaller difference in age between \M{3}\ and
\M{13} --- should that turn out to be the case --- unless something other
than, or in addition to, age is varying (see Catelan and
de~Freitas~Pacheco 1995). On the other hand, VBS found {\it no\/}
detectable age range in excess of some 0.3~Gyr ($\sim$2\%) among the most
metal-poor clusters with the best photometry; these particular results
thus tend to favor ELS, especially the Sandage (1990) version.

Therefore, the evidence from nearby clusters suggests that understanding
the formation of the Galactic halo is not so much a matter of deciding
which of ELS and SZ is ``right,'' but rather accepting that there is some
truth in both pictures, and trying to decide the fraction of the halo that
is best described by each of them.  About six orders of magnitude separate
the mass of the typical globular cluster from the mass of the Galaxy.
Where within this range did the mass of the typical cluster-forming lump
lie?  What was the ratio of kinetic to gravitational potential energy in
the protoGalaxy at the instant the local Hubble expansion ceased and the
contraction began?  Before we undertake to answer these questions, it must
be stressed that the observational studies discussed so far have had
little to say about clusters far from the Solar circle, and thus are
almost certainly unfairly sampling the full range of circumstances that
prevailed in the protoGalactic halo.  To obtain the big picture, it will be
necessary to measure precise relative ages for clusters sampling different
extremes of ancestry:  those close in to the Galactic center, and those
at the outermost fringes of the modern-day halo.  Reddening, crowding, and
confusion with field stars complicate observations of the former, while
sheer distance makes the latter difficult to study in any detail.  But,
before progressing to a discussion of the most recent age determinations
for globular clusters, it is worthwhile making a slight diversion to a
discussion of how relative ages can be measured.

\bigskip

\centerline{4.\ DIFFERENTIAL AGE-DATING TECHNIQUES}
\nobreak
Since the two dominant competing models for the formation of the Galaxy
make conflicting predictions for the relative duration of the
halo-formation epoch (\ie $\sim$1\% in classical ELS or $\lta$10\% in
Sandage's (1990) more recent reformulation, as compared to $\sim$30\% in
classical SZ and 30--100\% in Zinn's (1993) most recent extension of that
model), precise {\it relative\/} age determinations should suffice to
distinguish between the two extremes of possibility.  This is fortunate,
because relative age determinations can use stellar evolution theory in a
strictly differential sense, removing most of the effects of theoretical
uncertainties in absolute chemical abundance ratios, opacities, convection
formalism, temperature-color relations, and the like.  Differential
comparisons can also be devised which reduce the effects of observational
uncertainties in the absolute distance scale, overall metal abundance, and
individual cluster reddenings.

The differential age-dating techniques for globular clusters in use at the
present time can be separated into two basic classes:  those that are
``vertical'' and those that are ``horizontal'' in the color-magnitude
diagram.  Each has its advantages and its problems.  The most venerable of
the vertical techniques is the magnitude difference between the luminosity
of the horizontal branch --- as measured from the RR Lyrae stars or the
nonvariable stars bordering the instability strip --- and the
main-sequence turnoff, \delv, as illustrated in Figure~1.  From the very
first theoretical models of horizontal-branch stars having helium-burning
cores and hydrogen-burning shells (Faulkner 1966), it has been known that
the luminosity of the horizontal branch is only a very weak function of
total stellar mass and hence of age, with the main effect of increasing
total mass being to shift the star to the red along the zero-age
horizontal branch (ZAHB).  The mass of the helium core, which has some
influence on the luminosity of the ZAHB, is believed to be only a very
weak function of the total stellar mass, and hence age.  Therefore, the
horizontal branch should be a good magnitude fiducial from which to
measure the luminosity of the main-sequence turnoff, which fades at a rate
of some 0.01~mag for each 1\% increase in the age (\eg Green, Demarque,
and King 1987; Bergbusch and VandenBerg 1992).

Surveys of GC ages using this vertical technique in a strictly {\it
differential\/} sense --- the so-called ``$\Delta V$'' method --- were
carried out by Gratton (1985) and Peterson (1987), who came to conflicting
conclusions:  Gratton found from a sample of 26 clusters that the age
spread in the inner halo was negligible, $\lta$1.5~Gyr, with a mean age of
some 15~Gyr (\ie an age range $\lta$10\%), while that in the outer halo
was larger, at least 20--30\%, in essential agreement with the SZ model.
Peterson, on the other hand, concluded that the age spread in his sample
of 41 clusters could not be distinguished from the observational noise
inherent to the technique, which amounted to $\pm 17$\% or so.  Peterson
attributed the differences between his conclusions and Gratton's to three
causes.  First, his own sample was based on a complete survey of the
literature, while Gratton's was selected from a relatively small set of
papers; as a result, different numerical values of \delv\ were adopted for
some clusters in the two surveys.  Second, Peterson adopted a slightly
different metallicity scale for the most metal-rich clusters, noting,
however, that by itself this difference would account for no more than 3\%
in age.  Finally, Gratton had calibrated the absolute magnitudes of the
horizontal branch as a function of metallicity in two ways:  (a) on the
basis of theoretical stellar models for HB stars (Caloi, Castellani, and
Tornamb\`e 1978), ${d\,M_V/d\,\hbox{\rm [Fe/H]}} = +0.16$, and (b) on the
basis of the observed period-luminosity-amplitude relation for cluster
RR~Lyraes (Sandage, Katem, and Sandage 1981; Sandage 1982),
${d\,M_V/d\,\hbox{\rm [Fe/H]}} = +0.315$.  Gratton expressed a preference
for the latter because it produced a greater consistency among the ages
for clusters in the inner halo (in a sense, assuming that which he was
trying to determine).  Peterson rejected the latter metallicity-magnitude
relation because it had little independent observational or theoretical
support, but noted that this difference tended to affect the absolute age
scale rather than the variation of age with Galactocentric distance.  In
addition to the points raised by Peterson, it should also be remarked that
Gratton did not consider the observational errors in the individual age
determinations, most particularly that the data for the more remote
outer-halo clusters might simply contain larger random errors than for the
closer inner-halo clusters.  Conversely, it is also possible that
Peterson's more heterogeneous sample included some poorer data, thus
helping the observational noise to swamp the signal in his study.
However, Peterson did reexamine his conclusions on the basis of the 17
clusters with the best observations, and found no significant difference
from the results based on his full sample.

More recently, Sarajedini and King (1989) used the difference between HB
and turnoff magnitudes to estimate the ages of 31 Galactic GCs.  They
considered three different metallicity-luminosity relations for the
horizontal branch:  (a) the ``classical'' constant-luminosity assumption,
${d\,M_V/d\,\hbox{\rm [Fe/H]}} \equiv 0$; (b) a relation based on
theoretical HB models (Lee, Demarque, and Zinn, private communication ---
presumably those used by Lee \etal\ 1994 --- with a zero-point adjustment
to correct from $Y=0.23$ to $Y=0.24$), which was also roughly consistent
with observed luminosities based on the Baade-Wesselink method,
${d\,M_V/d\,\hbox{\rm [Fe/H]}} = +0.17$; and (c) Sandage's (1982, and
references therein) period-luminosity-amplitude relation,
${d\,M_V/d\,\hbox{\rm [Fe/H]}} = +0.35$.  As with Gratton's (1985) study,
both the absolute ages and the slope of the age--metallicity relation were
found to be sensitive to the choice of metallicity--luminosity law for the
horizontal branch, with luminosity independent of metal abundance implying
a strong age-metallicity relation and Sandage's relation implying none.
All HB relations implied an age distribution with a full-width at
half-maximum of roughly 45\%, implying a one-sigma dispersion of 19\%, and
a total range in excess of 60\%, resembling Zinn's (1993) ongoing
accretion model even more closely than SZ's slow contraction
model\footnote{$^2$}{However, it is possible that the inferred full-width
at half-maximum has been inflated somewhat by the authors' use of the
``generalized histogram'' to characterize the distribution of cluster ages
(their Fig.~1).  The age distribution of the Galactic globular clusters
has some true, intrinsic width.  The observed distribution will be
somewhat broader than the true one, due to random observational errors.
In creating the generalized histogram the observed distribution is
convolved with a Gaussian kernel having a standard deviation of the same
order of magnitude as the observational errors.  The generalized histogram
has thus been broadened by the error distribution twice, once by Nature
and once by the astronomer.  A perhaps more valid estimate of the true
intrinsic age dispersion would be obtained by {\it subtracting\/} the
variance of the observational errors from the sample variance; in this
case, the intrinsic age dispersion of the halo clusters would be reduced
from 19\% to 8\% (1.1~Gyr for a mean age of 14.3~Gyr) if Sarajedini and
King have correctly estimated the observational errors.}.

Finally, the $\Delta V$ method has very recently been applied to 43
Galactic globular clusters by Chaboyer, Demarque, and Sarajedini (1996; CDS).
Adopting a range of slopes ${d\,M_V/d\,\hbox{\rm [Fe/H]}}$, they derive
age ranges for the Galactic globular cluster population.  For a reasonable
intermediate metallicity slope, ${d\,M_V/d\,\hbox{\rm [Fe/H]}} = +0.20$,
they infer an age range ($\pm$2$\sigma$) of some 9~Gyr for the Galactic
globular clusters as a whole, with a {\it mean\/} cluster age of some
18~Gyr.  However, they recognize that four of the clusters in their sample
(Ter~7, Pal~12, IC$\,$4499, and \ngc{6652}) are considerably younger than
the rest and may represent a distinct population; removing these clusters
from the sample, they derive a reduced age range (again $\pm$2$\sigma$) of
$\sim$5~Gyr (\ie $\sigma = 1.25$~Gyr) for the remainder.  To arrive at
these numbers, they have subtracted (in quadrature) the estimated
observational errors in [Fe/H] and \delv\ from the observed dispersion in
the apparent ages; these are estimated to amount to $\pm$1.6~Gyr.  The
observational uncertainties in \delv, in particular, were calibrated by
comparing independent determinations of \delv\ for 13 clusters, from which
CDS concluded that the actual external errors of the published
magnitude differences were in fact only 0.61 times as large as the
original authors had estimated.  The conclusions of the paper rest heavily
on this assertion that numerous independent investigations have
systematically overestimated their observational errors by some 64\%:  the
observed dispersion in the cluster ages {\it including\/} the four
anomalously young ones is only $\sim$3~Gyr; it is the quadrature
difference between this and the typical estimated age uncertainty of
1.6~Gyr that leads to an inferred value of $\sim$2.25~Gyr for the
intrinsic age dispersion in the halo.  Had the uncertainties in
\delv\ given in the original papers been adopted instead, the uncertainty
in the individual ages would have been taken to be typically some 2.6~Gyr,
implying an age dispersion of $\sqrt{(3)^2-(2.6)^2} \approx 1.5$~Gyr {\it
including\/} the four anomalously young clusters, and allowing for no
intrinsic age dispersion at all if those four clusters are set aside.

And it is not clear whether this ``correction factor'' --- the ratio of
perceived external to internal error --- is well justified.  It is based
on multiple studies of a few well-observed, fiducial clusters that the
authors have tabulated and discussed in an Appendix.  Starting at the top
of their list, three citations are given for 47~Tucanae:  Hesser,
\etal\ (1987); Chaboyer, Sarajedini, and Demarque (1992); and Sarajedini
and King (1989; SK).  In going back to those sources, we find that the
\delv\ published by Chaboyer, Sarajedini, and Demarque is taken from
Buonanno, Corsi, and Fusi~Pecci (1989; BCF) who in turn state that their
value is based on the photographic data of Lee (1977), while SK state that
{\it their\/} value for \delv\ is itself based on the data of Hesser,
\etal\ (1987).  The second cluster on their list, \ngc{288}, has citations
to BCF, SK, and Bergbusch (1993) taken in conjunction with Pound, Janes,
and Heasley (1987).  Again going back to these sources, we find that BCF
cites previous photographic work by Buonanno and collaborators for the
horizontal-branch photometry, while the turnoff magnitude was determined
from their own CCD data.  SK list Olszewski, Canterna, and Harris (1984)
and --- again --- Pound \etal\ (1987) as the source of the data used in
their analysis.  It is difficult to understand how the mixing of
photographic and CCD data, and how the web of interlocking citations will
have affected the validity of this ``correction factor'' of 0.61.
Grundahl (1996) has compared the ``vertical'' ages of CDS with the
``horizontal'' (see below) ages of Richer \etal\ (1996).  He concluded,
``This comparison revealed large discrepancies between the two methods and
the likely source found to be [{\it sic\/}] due to errors in the adopted
\delv\ values from the literature $\ldots$\ \  It was also argued that
this indicates that different authors do not determine the location of the
TO in consistent ways and that one is most easily led to an
underestimation of the true values.''

Whatever one concludes about the reality of the intrinsic dispersion of
globular cluster ages, CDS found, as did Gratton and Sarajedini and King
before them, that for small values of ${d\,M_V/d\,\hbox{\rm [Fe/H]}}$,
there is a distinct tendency for more metal-rich clusters to be younger
than metal-poor ones.  CDS quantify this as there being a high probability
($> 98$\% confidence level) of metal-rich clusters being younger than
metal-poor ones for all values ${d\,M_V/d\,\hbox{\rm [Fe/H]}} < 0.24$
(their Fig.~2); however, according to them, this trend is superimposed
upon a large apparent age spread at each metallicity.  No compelling
evidence was found for a trend of age with Galactocentric distance,
although it must be noted that their sample did contain only one cluster
outside a Galactocentric radius of 22~kpc.

The ``horizontal'' method of measuring ages offers an alternative to the
vertical technique.  VBS and Sarajedini and Demarque (1990; SD) formalized
some folk wisdom that had been understood by the stellar evolution
community for decades: within limits, stellar evolution tracks and model
cluster isochrones are rather self-similar.  Differences in mass and
chemical abundance produce loci that differ in their absolute vertical and
horizontal position in the theoretical Hertzsprung-Russell diagram but,
in fact, their {\it shapes\/} do not differ by very much.  By shifting
theoretical isochrones both vertically and horizontally in the H-R diagram
in order to register them at the magnitude and color of the turnoff, VBS
demonstrated graphically that the principal morphological difference
between isochrones was a shrinking of the color difference between the
turnoff and the lower part of the giant branch with increasing age (see
their Fig.~2).  At least for the lower-metallicity clusters ([Fe/H]
$<-1.2$ or so), the effect of metal abundance appeared to be negligible
(see VBS's Fig.~3), although the authors did stress that it would be
dangerous to apply this technique to the comparison of clusters with
greatly different metallicities, because the errors in the stellar models
--- due, \eg to our imperfect understanding of convection and opacities
--- might themselves depend on $Z$.  Therefore, given a sample of
clusters of similar metal abundance, by simply measuring the differences
in color between the bluest point at the turnoff and the lower part of the
giant branch in some self-consistent way, one has a direct measure of the
spread in age within the sample.

VBS demonstrated that the most metal-poor of the nearby clusters with the
best photometry, \ngc{4590}\ = \M{68}, \ngc{6341}\ = \M{92}, \ngc{6397},
and \ngc{7099}\ = \M{30}, with [Fe/H] $\sim-2$, show no perceptible range
in age (Figure~2): the measured dispersion corresponds to $\sim$0.3~Gyr,
which is consistent with the measuring errors.  Furthermore, considering
this group of GCs, stellar samples taken from different parts of the same
cluster show variations like those found between different clusters, again
suggesting that these differences are consistent with the measuring
errors.  Subsequently, Heasley and Christian (1991) and Durrell and Harris
(1993) have reported that the ages of \ngc{5024}, \ngc{5053} and
\ngc{7078}\ = \M{15}, too, are indistinguishable from that of these four
clusters.  Conversely, among the most-metal rich of the clusters that they
considered ([Fe/H] $\sim$ --1.3), VBS concluded that \ngc{288}\ was
probably some 2.5$\pm$1.5~Gyr older than its second-parameter counterpart,
\ngc{362}, and most of the other clusters in that abundance range
(\ngc{1261}, \ngc{2808}, \ngc{3201}, \ngc{5904}\ = \M5, and Palomar~5).
The famous anomalous cluster Palomar~12 may lie just outside the abundance
range where the horizontal method may be applied with confidence, but when
the method is used nonetheless, an age difference of some 25\% between
Palomar~12 and more typical clusters is confirmed.  At intermediate
metallicities ([Fe/H] $\sim$ --1.6), the situation was rather less clear,
due primarily to a shortage of good published data.  Among most of the
clusters in this sample (\ngc{5272}\ = \M3, \ngc{6218}\ = \M{12},
\ngc{6752}, and \ngc{7492}), there was no compelling evidence for a range
of age.  The comparison of \M3\ with \ngc{6752}, in particular, suggested
that age is not responsible for the notable difference in the
horizontal-branch morphologies of these two clusters.  Conversely, a
comparison of \M3\ with some of Stetson's unpublished data for
\ngc{6205}\ = \M{13}\ suggested that the latter may be older than the
former by some 1--2~Gyr, which is of the correct sign but falling somewhat
short of the 3~Gyr difference predicted from their HB morphologies.  So
the conclusions of VBS were: first, that the age spread in the available
sample appeared to be unmeasureably small among the most metal-poor
clusters:  almost certainly $\lta$0.5~Gyr, or some 3\%, which would be a
remarkable coincidence if the true duration of the cluster-formation epoch
was of order 5~Gyr, or $\sim$40\%.  Second, among the
intermediate-metallicity clusters, there was a hint of the possibility ---
no more --- that age might (sometimes?) be a second parameter; \ie the
results were ambiguous.  Third, among the more metal-rich clusters, there
was stronger evidence for a range of age, up to as much as 25--30\% if
Pal~12 represents one extreme of the age range for normal clusters, or as
little as 2--3~Gyr (12--20\%) if Pal~12 is a true anomaly with an origin
quite different from that of the main halo cluster population.  SD applied
their own version of the horizontal method specifically to the
\ngc{288}--\ngc{362}\ comparison, and likewise concluded that \ngc{288} is
older (by some 2.5--3~Gyr).

Each of these approaches, the vertical and the horizontal, is doubly
differential: the difference between two things measured in one cluster is
compared to the same difference measured in another.  Both are independent
of distance, reddening, and errors in the assumed photometric zero
points.  Apart from this common strength, the advantages and disadvantages
of the two methods are different.  The vertical method is on a sounder
theoretical footing, because the luminosity of the turnoff depends
primarily on properties and processes in the stellar core, where the
physics is simple and comparatively well understood.  By contrast, the
horizontal method is affected by the stellar radii, and hence might be
sensitive to less-well understood envelope physics such as opacities and
convection:  if these were to differ from our understanding of them in
ways which vary from cluster to cluster, then our age estimates would be
affected.  However, Figure~3 shows the same data as are plotted in Fig.~2,
but compares them to a set of isochrones which are identical except for
the choice of mixing-length parameter:  $\alpha_{\hbox{\sevenrm MLT}} =
2.5$, as opposed to 1.5.  The comparison of Fig.~3 with Fig.~2
demonstrates yet again that, while the horizontal method cannot at present
be used to determine absolute ages, a systematic error in our knowledge of
convection theory at some particular metal abundance will not endanger
conclusions about the relative age {\it range\/} at that abundance, unless
there is cluster-to-cluster variation in some {\it other\/} parameter that
alters the convection mechanism.  Similarly, Figure~4 shows that the
horizontal method is not particularly sensitive to unrecognized
cluster-to-cluster differences in overall metallicity (for [Fe/H] $<$
--1.2 or so), oxygen- and other $\alpha$-element relative abundances, or
helium abundance, {\it unless\/} a secondary effect of these hypothetical
differences is to alter the physics of the envelope convection.  But, as
discussed in VandenBerg \etal\ (1996), it has become apparent that
perceived turnoff magnitudes may also be subject to the vagaries of
modelling stellar $\Teff$'s, and hence the vertical method may not be
completely immune to these hidden-parameter uncertainties either.

However, all such theoretical concerns aside, the apparent magnitude of
the bluest point at the turnoff can be quite hard to measure, because the
stellar locus is (by definition) vertical at that point.  Theoretical
models predict that the color of the main sequence in {\it B--V\/} will be
constant to $\leq$0.01~mag over a range of $\sim 0.6$~mag, and constant to
$\leq$0.02~mag over a range of nearly 0.9~mag.  (VBS have shown that these
numbers do not depend sensitively on age or metal abundance.)  Thus,
slight calibration errors, stochastic arrangements of stars in the
diagram, the inclusion of composite-light binaries or marginal blue
stragglers in the sample, or the preconceptions of the astronomer
hand-drawing the mean locus, can easily bias the estimate of the vertical
position of the bluest point by a meaningful amount.  Conversely, the
verticality of the turnoff makes it exceedingly easy to estimate the {\it
color\/} of the bluest point.  By effecting the {\it vertical\/}
registration of two cluster sequences not by the apparent magnitude of
that bluest point, but by the magnitude of the point on the main sequence
at some fixed $\Delta$(color) redder than the bluest point, where the
slope of the cluster locus is appreciable, the VBS approach greatly
reduces the uncertainty in the vertical offset which, since the turnoff is
vertical and the lower RGB is nearly so, is of significantly reduced
importance in any case.

The vertical method relies on a fine measurement of the apparent
magnitudes of objects that differ in luminosity by 3--3.5~mag
(cf.\ Fig.~1), a factor of order 20.  Early on, many observing programs
were designed to do one of two things, but not both: either to define the
horizontal branch and the giant branch down to some level below the HB, in
which case the turnoff stars were undetected, or to measure the turnoff
and as far down the main sequence as possible, in which case fields were
chosen with a minimum number of giants and HB stars, and/or those brighter
stars were saturated.  Many of the earlier values of $\Delta V$ therefore
were based on combining two different data sets, with the shallower data
often coming from photographic work.  Programs intended to bracket the HB
and the turnoff within a single data set have more recently been
undertaken, but it still requires a major effort to obtain tractable
samples of {\it both\/} turnoff and horizontal-branch stars in the same
images, and such studies are still vulnerable to non-linearities in the
detector and to possible subtle differences in the way the data analysis
software treats bright, well-exposed stars as compared to faint, noisy
ones.  (For instance, in synthetic aperture photometry, bright stars are
optimally measured in large apertures, while faint stars are better
measured in small apertures, with an empirical correction to
large-aperture values (\eg Stetson 1989).  In profile-fitting photometry,
the dominant source of uncertainty in the measurement of bright stars is
point-spread-function mismatch; the measurement error of faint stars is
dominated by photon noise and the estimation of the local sky
brightness.)  In the horizontal method, the turnoff stars are compared to
stars that are both much more like them in apparent magnitude
(\cf.\ Fig.~1) and considerably more common than HB stars.  Furthermore,
the comparison is in {\it color\/} and any residual systematic
nonlinearities in either detector or analysis will tend to cancel when the
magnitudes are differenced to form the instrumental color index.

The purely observational uncertainties in the horizontal method can be
reduced simply by using colors with a longer wavelength baseline (Stetson
1993).  For instance, {\it B--I\/} is roughly 2.5 times more sensitive to
temperature than {\it B--V\/}, which produces the straightforward factor
of 2.5 improvement on the leverage for estimating temperature differences
given fixed photometric errors by, in effect, increasing the curvature of
the turnoff by a factor 2.5, and increasing the relative displacement of
the lower RGB from the turnoff by the same factor.  The use of {\it
B--I\/} also greatly aids in distinguishing the true main sequence from
binaries and marginal blue stragglers.  Furthermore, in {\it B--I\/} the
main sequence is more horizontal and the giant branch is more vertical
than in {\it B--V}, thus rendering the exact vertical registration of two
cluster sequences even easier and even less important.   There is no
analogous way to increase the power of the vertical method, since the
turnoff and the flat part of the horizontal branch have similar
temperatures, so the magnitude difference is roughly constant at all
wavelengths.  The similar colors of turnoff and HB stars free the vertical
method from errors in the color transformation from instrumetal to
standard photometric indices.  Still, with a color difference of order
0.25~mag (in {\it B--V\/}) between the turnoff and the base of the giant
branch, the horizontal method is not overly sensitive to transformation
errors of a percent or so, particularly when used to intercompare clusters
observed during the same observing run.

A final point about the vertical method is that it breaks down for ages
that are too small.  This is because, for horizontal-branch masses greater
than 0.9--1.0~${{\cal M}_\odot}$, the ZAHB turns upward and then returns
toward the blue, brightening all the while (\eg Faulkner 1966, Demarque
and Hirshfeld 1975).  The correctness of the models in this prediction is
dramatically illustrated by the CMD of the dwarf spheroidal galaxy in
Carina published by Smecker-Hane \etal\ (1994; their Figs.~2 and 3), where
the clump of core helium-burning stars belonging to a $\sim$6~Gyr
population lies roughly a quarter-magnitude above the horizontal branch of
a $\sim$15~Gyr population.  Therefore, below some age (which is greater
than 6~Gyr), the vertical method becomes ambiguous, as a given $\Delta V$
can correspond to two possible ages.  There is no corresponding barrier to
the use of the horizontal method at younger ages.  On the other hand, VBS
stress that their horizontal method cannot be applied with confidence
to clusters more metal-rich than some minimum value lying in the
neighborhood of $-1.2$:  for more metal-rich clusters the color difference
between the turnoff and the base of the giant branch depends upon both
metallicity and age.  Perhaps the horizontal method can still be applied
to determine differential ages for metal-rich clusters shown
spectroscopically to have closely similar abundances.  Alternatively,
Buonanno \etal\ (1993) have demonstrated that enforcing {\it
consistency\/} between vertical and horizontal techniques can
simultaneously constrain age and metallicity differences, at least toward
the metal-poor end of the globular-cluster abundance distribution; perhaps
the same sort of approach can be made to work for more metal-rich
clusters.  However, the propagation of accumulated random errors may limit
the ultimate precision achievable with such compound, multiply
differential techniques.

Other types of vertical approaches are possible.  For instance, given the
theoretical prediction that the luminosity at the helium flash is nearly
independent of the mass, and hence the age, of a star --- certainly for
any mass within the range $0.7\le {\cal M}/{{\cal M}_\odot} \le 1.0$
(cf.\ Sweigart and Gross 1978) --- the magnitude difference between this
reference luminosity and that of the turnoff can be used to determine
relative cluster ages.  Of course, the brightest giant in a globular
cluster will not be precisely at the RGB tip, but Monte Carlo simulations
by, \eg Crocker and Rood (1984) have shown that with as few as 45 stars in
the upper 2.5 mag of the giant branch, the brightest should be within 0.11
mag of the helium flash luminosity 68\% of the time.  (For this reason,
the helium-flash luminosity cannot be used as a reference in the Galactic
globular clusters of the lowest mass, because some of them do not contain
as many as 45 giants.)  Making use of the extensive surveys of the bright
giants in a large sample of GCs that had already been carried out (Frogel,
Persson, and Cohen 1981, 1983), VandenBerg and Durrell (1990) vertically
shifted the CMDs for several globulars having similar metallicities until
the brightest giant in each cluster had the same $V$ magnitude, and then
compared their respective turnoff magnitudes.  Those for \ngc{288},
\ngc{362}, and \M{5} were found to be sufficiently similar that VandenBerg
and Durrell concluded that the ages of these three cluster were ``not
detectably different" (i.e., any variation was less than the estimated
1--2 Gyr $1\sigma$ uncertainty).  The same conclusion was reached for the
group of very metal-poor GCs comprising \M{15}, \M{30}, \M{68}, and
\M{92}.  (It is encouraging that these results are close to those obtained
subsequently by VBS, who used a technique that is capable of much higher
precision.)

But this approach has many practical difficulties.  For one, as already
implied, the brightest giants are very rare, and the method is therefore
sensitive to small number statistics.  For another, the $V$ magnitude of
the RGB tip is a fairly strong function of the metal abundance (see, \eg
Fig.~1; also VandenBerg 1992), so that even relatively small differences
in chemical composition could, if not taken into account, lead to
misleading results.  It would be far better to use infrared photometry:
for instance, Da Costa and Armandroff (1990) have proposed that the
RGB-tip $I$-band luminosity is constant (or very nearly so) in GCs.  In
addition, giant-branch tip stars are far more luminous than even
horizontal-branch stars and, furthermore, they are extremely red: the use
of a long-wavelength filter (such as $I$) would further amplify the
magnitude difference between the tip stars and the much bluer turnoff,
greatly compounding the dynamic-range problem with the vertical method.

Fusi~Pecci \etal\ (1990), and subsequently Sarajedini and Lederman (1991),
have proposed that the magnitude of the ``bump'' in the giant-branch
luminosity function can be used to estimate GC distances.  It would follow
that the magnitude difference between the bump and the main-sequence
turnoff could also be used as a distance-independent age indicator.  The
practical difficulties with this method, too, would be large.  Apart from
the fact that the observed magnitude of the bump is not yet successfully
predicted by theory, in most cases it would be necessary to survey a major
fraction of a given cluster to obtain a sample large enough to define the
bump accurately.  Much of this area would be too crowded to measure the
turnoff stars with adequate precision, so the sample of ``bump'' stars and
of turnoff stars would be measured under significantly different
conditions.  Even then, unlike the horizontal branch, the bump differs
from the turnoff in color as well as magnitude, so the age index would be
sensitive to erors in the color transformation.  Anyway, with the current
generation of detectors, these magnitude fiducials will more probably be
used in the same way as Sandage (1953) used the horizontal branch, as a
means to estimate the absolute distance to a cluster, from which the
apparent magnitude of the main-sequence turnoff as determined by a
different set of observations can be used to infer the age, an approach
which includes the difficulty of making sure the two studies are truly
calibrated to the same system and, hence, is not purely differential.

\bigskip

\centerline{5.\ A TEST OF THE HYPOTHESIS THAT AGE IS THE SECOND
PARAMETER}
\nobreak
In their latest paper, Lee, Demarque, and Zinn (1994) presented a strong
case that age must be the cause of the observed variations in HB
morphology among clusters of the same [Fe/H] (the ``first parameter'') and
as a function of Galactocentric distance ($R_G$).  They did so by the
process of elimination; that is, they demonstrated that variations in any
of the other usual second-parameter suspects --- helium abundance, CNO
abundance, and core mass (Rood 1973, Renzini 1977) --- lead to
inconsistencies with observations.  Furthermore, they argued that the
recently discovered correlations of the length of the blue HB tail with
cluster density and concentration (Fusi Pecci \etal\ 1992, Buonanno 1993),
which might be suggesting that environmental influences drive differential
mass loss, is not apparent in the observed trend of HB type with $R_G$.
Only the age explanation appeared to be comparatively trouble-free, and in
view of the indisputable fact that young GCs {\it do\/} exist in the
Galaxy, the evidence seemed compelling to them that age
is the dominant (possibly the sole) second parameter.

Figure~5 shows a composite version of one of the diagrams contained in the
Lee \etal\ (1994) paper (their Fig.~7).  It plots the HB type, as
quantified by $(B-R)/(B+V+R)$ --- where $B$, $V$, and $R$ represent the
numbers of blue HB stars, RR Lyrae variables, and red HB stars,
respectively --- as a function of [Fe/H] for 83 GCs.  Different symbols
are used to indicate the division of the clusters into three radial
zones:  the solid lines illustrate the Lee \etal\ synthetic HB
calculations assuming that all stars undergo the same amount of mass loss
during the giant-branch evolution and that only [Fe/H] varies from cluster
to cluster.  The three curves are effectively ``HB isochrones,'' with the
upper/lower ones showing how the position of the middle locus would be
changed by an age increase/decrease of 2 Gyr.  This is admittedly a very
tantalizing figure, showing as it does how the spread in HB type at a
given [Fe/H] and the variation in HB morphology with $R_G$ can be
accounted for by differences in age.

However, as already noted, there are some difficulties for the pure age
hypothesis.  Using their $\Delta$(color) technique, VBS derived
differences in age between \ngc{288} and \ngc{362} and between \M{3} and
\M{13} that are smaller than the ones implied by Fig.~5.  According to the
``HB isochrones'' computed by Catelan and de~Freitas~Pacheco (1993), for
any reasonable estimate of the absolute cluster ages, \ngc{288} must be
$\gta 5$--6 Gyr older than \ngc{362}, if age alone is responsible for the
differences in their HB morphologies.  Lee (1991) has claimed that he
could reduce the age difference to $\sim$3--4~Gyr if the stars in both
clusters had [O/Fe]~=~0.85, but such a high oxygen abundance has pretty
well been ruled out observationally:  Dickens \etal (1991) have obtained
[O/Fe]$\approx$+0.2 in unmixed stars in both \ngc{288} and \ngc{362};
Sneden \etal\ (1991, 1992, 1994) and Kraft \etal\ (1992) have obtained
similar values in clusters both more metal-rich and more metal-poor than
these; Bell, Briley, and Norris (1992) have argued that CO band strengths
in the metal-poor GC \ngc{6397} are incompatible with O abundances as high
as [O/Fe] = +0.6; and Brown and Wallerstein suggest that [O/Fe] $\approx$
+0.5 over the entire range of iron abundances found in $\omega$~Centauri.
Lee has also remarked that a large difference in [Fe/H] between the two
clusters could accomplish the same thing, but this suggestion appears to
be untenable as well (see, \eg the spectroscopic studies by Caldwell and
Dickens 1988 and Dickens \etal\ 1991).  In addition, based on the $\Delta
V$ method, Stetson \etal\ (1989) have inferred that \ngc{362} and \M{5}
have very similar ages:  the same conclusion, using the same approach, was
reached by Catelan and de~Freitas~Pacheco (1995) for the \M{3}--\M{13}
pair.  These results clearly conflict with the expectations from Fig.~5,
as does the recent work by Stetson and VandenBerg (1996) on \M{2} and
\M{3}.  The latter have found, from differential photometry obtained on
the same night with the same telescope, that the main-sequence to
lower-RGB CMDs of these two clusters, which also form a second-parameter
pair (see Fig.~5), are virtually identical, implying indistinguishable
ages.

There are at least three further problems for the age interpretation of
the variations in HB type among GCs.  First, taken at face value, Fig.~5
suggests that all of the clusters with extremely blue HBs, including
especially the intermediate-metallicity cluster \ngc{288}, are much older
than {\it all\/} of the extremely metal-deficient systems like \M{15} and
\M{92}.  This seems highly improbable, as also noted by
Bergbusch (1993), who found that isochrone fits favored very similar ages
for \ngc{288} and \M{92}.  Second, among the ``young'' globular clusters,
Arp~2 has a ``normal-looking'' {\it blue\/} HB for its metallicity, in
spite of its relative youth (see Buonanno \etal\ 1994).  The same comment
could be made concerning IC~4499 (Ferraro \etal\ 1995), whose HB looks
very similar to that in \M{3} and which has a similar metallicity, but is
certainly younger.  Some additional parameter must be at work which is
compensating for the effects of younger age on the HB morphologies of
these two clusters.  And third, the pure age explanation cannot be
reconciled with the observations of bimodal-HB clusters like NGC 1851
(Walker 1992) and NGC 2808 (\eg Rood \etal\ 1993), which possess both
very blue and very red HB populations with few stars between them.

This brings us to our simple test of the age hypothesis: let us use
the CMD for a bimodal-HB globular cluster as a template with which to
compare those of \ngc{288} and \ngc{362}.  To be specific, let us force a
coincidence of the turnoff luminosities of all three clusters, to be
consistent with the working hypothesis that they are coeval, and then
intercompare their bright-star populations to check the validity of that
assumption.  We have chosen to use the \ngc{1851} CMD as obtained by
Walker (1992) to act as this bridge because its photometric sequences
are much better defined than existing data for \ngc{2808}, which has a
similar metallicity.

Figure~6 shows the superposition of the upper main sequences, the subgiant
branches, and the lower RGBs of \ngc{288}, \ngc{362}, and \ngc{1851} when
the horizontal and vertical offsets listed in the caption are adopted.
There is clearly some uncertainty in the choice of these offsets, given
the obvious differences in the cluster fiducials, but by making the
age-dependent subgiant branches coincide, we remove most (if not all) of
the ambiguity inherent in the matching of turnoff luminosities alone.
Figure~7 illustrates the upper part of the \ngc{1851} CMD on which
hand-drawn envelopes to the stellar distributions are indicated by the
solid curves.  Then, in Figure~8, these same hand-drawn boundaries are
superimposed onto the bright-star populations of \ngc{288} and \ngc{362}
using those offsets that were adopted in producing Fig.~6.

The agreement is remarkable: \ngc{1851} contains a red HB just like the
one in \ngc{362} {\it and\/} a blue HB very much like that in \ngc{288}.
(Although the stars at the top end of the blue tail in \ngc{288} appear to
be somewhat brighter than those in \ngc{1851}, this could well be expected
if the former evolved from fainter ZAHB locations than the latter, which
would be consistent with the observation that \ngc{288} has a longer blue
tail than \ngc{1851}; cf.\ Fig.~1.)  The main implication of this
comparison is that all three clusters do, indeed, have the same age to
within quite a small uncertainty ($\lta\pm 1$ Gyr).  Moreover, small
cluster-to-cluster differences in [Fe/H] or [$\alpha$/Fe] will not alter
this conclusion because we have effectively used the $\Delta V$ method,
which is insensitive to modest changes in heavy-element abundances (see,
\eg Bencivenni \etal\ 1991, Caputo and Degl'Innocenti 1995).  In addition,
since \ngc{1851} {\it contains\/} the \ngc{288} and \ngc{362} CMDs,
differences in $Y$ or [CNO/Fe] cannot be the explanation of the diversity
in HB morphology among GCs either, unless such differences occur among the
stars of this one cluster (which seems unlikely).

The \ngc{288}--\ngc{362}--\ngc{1851} comparison is perhaps the strongest
evidence so far --- and there have been many other indications (some noted
above) --- that HB-morphology arguments alone (cf.\ SZ, Lee 1992, Lee
\etal\ 1994) cannot be used with confidence to support a particular
chronology for the formation of the Galaxy.  It also shows how risky it is
to apply the $\Delta V$ method to those systems that have no zero-age HB
stars at the color of the instability strip.  For instance, Sarajedini and
King (1989) (also see Chaboyer and Demarque 1994) inferred that \ngc{288}
was much older than \ngc{362} because, in their estimation, the magnitude
difference between the HB and the turnoff was 0.33 mag larger in the
former cluster than in the latter.  On the other hand, it is not certain
that the extension of the ZAHB appropriate for the red HB stars in
\ngc{1851} would coincide exactly with the ZAHB from which the blue HB
stars evolved.  For instance, if core rotation were more important in the
bluer stars, then they may have started their core helium-burning phases
with slightly higher core masses than those to the red of the instability
strip.  But the difference would be small: the reddest of the blue HB
stars in \ngc{288} must have evolved from much bluer and fainter ZAHB
positions, as is suggested by the HB evolutionary tracks shown in Fig.~1
above.  (A similar conclusion is drawn if comparison is made with
\ngc{2808}, although these data are poorer.)

But why have color-difference methods (VBS, SD) also found \ngc{288} to be
older than \ngc{362}?  There may be a number of reasons.  It is still
possible that the two clusters do differ in age by $\sim$1--2~Gyr given
the inaccuracies in the available photometry (see Sandquist et al., 1996
for a recent comparison of \ngc{288}, \ngc{362} and \M5).  Only with
excellent photometry would the color-difference method be able to measure
an age difference to this level of precision.  In addition, the previous
applications of the color-difference method to these GCs were limited by
the small number of evolved stars that had been observed (a fact sometimes
masked by the use of fiducial lines in comparison plots).  Small-number
statistics in conjunction with the known observational imprecision may
well have dominated the $\Delta(color)$ measurements (recall that the
formal error in the VBS determination of the age difference between
\ngc{288} and\ngc{362} was $\pm 1.5$ Gyr).  In this day of large-format
CCDs, it would be relatively straightforward to improve the samples in
both clusters by a large factor.  Finally, \ngc{288} is known to have a
significant population of main-sequence binaries (Bolte 1992), as
indicated by the strong redward asymmetry in the distribution of stars on
the main sequence.  This could cause the estimated turnoff color to be
redder than that of the single-star turnoff, thereby simulating an older
cluster.

We can only speculate about what {\it else\/} could be causing the
variation of HB morphologies within the bimodal HB clusters, and from
cluster-to-cluster at a given [Fe/H] and as a function of $R_G$.  Core
rotation may be an important factor, though it would be hard to explain
the dependence of the second-parameter effect on position in the Galaxy in
terms of angular momentum differences (see Norris 1981, 1983).  The
concentration of a cluster or its central stellar density may also play a
key role in this phenomenon (Fusi Pecci \etal\ 1992, and references
therein).  As noted by Renzini (1983), nearly all clusters with gaps on
the HB have high central concentrations: close encounters may lead to the
stripping of stellar envelopes or to the spinning-up of significant
numbers of stars.  Possibly increased angular momentum is a natural
consequence of high concentration (Buonanno, Corsi, and Fusi~Pecci 1985;
Buonanno \etal\ 1986): according to Djorgovski and Meylan (1994), GCs
closer to the Galactic center tend to be more concentrated with smaller,
denser cores.  But these ideas can hardly apply to \ngc{288}, which is a
low-concentration GC, though gaps do seem to exist on its horizontal
branch as well (see Fig.~8).  However, perhaps the large binary population
in this cluster (Bolte 1992, Bolte and Dubath 1996) is the critical clue
to what is going on.  The way in which the protoGalaxy collapsed may also
have given rise to second-parameter effects (van den Bergh 1993).  Indeed,
it may well be that all of these contribute to a greater or lesser
extent.

The bimodal-HB clusters like \ngc{1851} and \ngc{2808} are unquestionably
very important Rosetta stones and they should be exhaustively studied.
One certainly has the impression from Figs.~7 and 8 that whatever is
afflicting \ngc{1851} is also the cause of the difference in the HB
morphologies of \ngc{288} and \ngc{362}.

\bigskip

\centerline{6.\ RECENT DEVELOPMENTS} 
\nobreak 
The launch and subsequent repair of the {\it Hubble Space Telescope\/} have
finally permitted the determination of precise ages for globular clusters
deep in the Galactic bulge and out at the extremes of the halo.  Using the
Planetary Camera of the original WF/PC on the unrepaired {\it HST\/},
Fullton \etal\ (1995) have obtained a color-magnitude diagram in the
$(V,I)$ bandpasses for the globular cluster \ngc{6352}\ which, according to
the dichotomy discovered by Mayall (1946) and Kinman (1959), and
subsequently confirmed by Zinn (1985), belongs to the disk subsystem of
globulars.  By applying the $\Delta V$ method to \ngc{6352}\ with respect
to 47~Tucanae (also a cluster of the disk subsystem, but slightly more
metal-poor and therefore requiring a small adjustment to $\Delta V$ to
allow for the abundance difference), they concluded that \ngc{6352}\ was
the older of the two by 0.7$\pm$2.2~Gyr.  Ortolani \etal\ (1995) present
analysis of data obtained with WFPC2 of the near-Solar metallicity bulge
clusters \ngc{6528} and \ngc{6553}. Based on either the horizontal or
vertical differential age indicators, these two clusters appear to be
identical in age to $\lta 10$\%. Comparing the clusters' luminosity
functions to that measured for the bulge field in one of Baade's Windows,
after registering the position of the HB, they find a very close
correspondence of the cluster and field main sequences indicating nearly
identical ages for these clusters and the field stars of the bulge. The
authors also make a case that the clusters' ages are close to that of
\ngc{104} (47~Tucanae), which is down in metallicity by a factor of 4--5
compared to \ngc{6528} and \ngc{6553}.  Because the effects of metallicity
are very large at the metal-rich end of the scale, this conclusion is less
easily justified.  With that caveat in mind, it nevertheless appears that
there is no evidence for a significant age range within this sample of
objects from the inner Galaxy.  Finally, recent work by Grundahl (1996) led
to the conclusion that the moderate-metallicity thick-disk GC's M71,
47~Tuc, \ngc{6352}, and \ngc{6760} have ages that are indistinguishable at
the level of 1.5~Gyr, suggesting that the oldest part of the thick disk
formed over a very short time span.  Conversely, the near-Solar metallicity
clusters \ngc{5927}, \ngc{6528}, \ngc{6553} formed some 3--4~Gyr later;
following Burkert, Truran, and Hensler (1992), Grundahl suggests that these
clusters may have formed along with the first generation of {\it
thin\/}-disk stars.

Stetson \etal\ (in preparation) have used the Wide Field and Planetary
Camera 2 on the post-repair {\it HST} to derive color-magnitude diagrams
for the remote-halo clusters \ngc{2419}, Pal~4, and Pal~3.  All three have
the large core radii typical of the diffuse globular clusters of the outer
halo, but while Pal~4 and Pal~3 both also have the very low masses
characteristic of outer halo clusters, \ngc{2419}\ is the third most
luminous Galactic globular cluster known.  In addition, while Pal~3 and
Pal~4 share the anomalously red horizontal branch implying the operation
of the second parameter, the horizontal branch of \ngc{2419}\ is blue
(normal for an inner-halo cluster with its measured metal abundance,
[Fe/H] $\approx -2$).  The vertical and horizontal methods as applied to
these clusters demonstrate that the age of \ngc{2419}\ is
indistinguishable at the level of 0.5--1~Gyr from that of \ngc{6341}\ =
\M{92}, a cluster with similar metal abundance and HB morphology, while
the age of Pal~3 is indistinguishable at a similar level of precision from
that of \M5, a cluster with comparable metal abundance but having a very
much bluer horizontal branch.  The vertical and horizontal methods as
applied to Pal~4 compared with \M{5}\ are somewhat ambiguous, in the sense
that both appear to suggest that Pal~4 is younger by several Gyr, but
there is not a good correspondence of mean cluster loci at all
magnitudes:  whether this anomaly is real or due to some error in the
photometric analysis is still under investigation.  But the first hints
from these studies are that (1)~age is not the sole cause of the diversity
in HB morphologies, and (2)~the variation of mean age with $R_G$ is small
and possibly even zero.

\bigskip

\centerline{7.\ SUMMARY} 
\nobreak
As the result of the superb color-magnitude data that have been obtained
through the use of CCD detectors on large telescopes and the {\it Hubble
Space Telescope\/}, relative GC ages are starting to be much better
known.  The application of color- and magnitude-difference techniques to
the observed CMDs is beginning to suggest that there is little or no
dispersion in age, either at a given [Fe/H] or as a function of
Galactocentric distance ($R_G$).  (There are undeniably a number of young
GCs in the Galaxy, but they appear to represent a small and quite distinct
minority and may well have been tidally pulled out of the Magellanic
Clouds and/or a nearby dwarf spheroidal.) In this study, we have provided
fairly strong (supporting) evidence that age is not the sole cause of the
second-parameter phenomenon, if indeed it ever plays an important r\^ole
at all.  Consequently, the observed changes in horizontal-branch type with
$R_G$ can probably not be explained in terms of age variations alone.

Whether or not GC ages vary with [Fe/H] is almost certainly the hardest of
the relative-age questions to answer.  To do so will require, among other
things, the resolution of present distance-scale uncertainties (mainly
concerning the variation of horizontal-branch luminosities with metal
abundance), and an improved understanding of the chemistry of stars
(notably the variation of [$\alpha$/Fe], and especially [O/Fe], with
[Fe/H]).  We suspect that the confluence of various streams of thought,
including but not limited to parallaxes and proper motions from HIPPARCOS,
MACHO-spinoff studies of variable stars in the Large Magellanic Cloud and
the Galactic bulge, {\it HST\/} observations of globular clusters in the
Local Group, and advances in stellar evolution theory, will provide us
with a trustworthy resolution of this question relatively soon.  

These matters aside, the recent results discussed in the previous
sections suggest that the Eggen, Lynden-Bell, and Sandage (1962) picture
of the formation of the Galaxy (especially the updated version described
by Sandage 1990) seems to be a good description of the dominant formation
mode of the Galactic GC system that we observe today, but in addition
there exists ample evidence that accretion/merger processes such as those
envisioned by Searle and Zinn (1978) and Zinn (1993) have also contributed
to the Galactic halo.  We hasten to add, however, that no clear consensus
has yet emerged --- this remark represents only {\it our\/} interpretation
of the available evidence as of mid-April, 1996.

\vskip 1.0 true in 

\noindent We are very grateful to Peter Bergbusch, Young-Wook Lee, and Alistair
Walker for providing, in machine-readable form, the data that have been
plotted in Figures 8, 5, and 7, respectively.  We also thank M\'arcio Catelan
for helpful discussions and Jim Hesser for his careful reading of the
manuscript.  D.A.V. acknowledges, with gratitude, the award of a Killam
Research Fellowship from The Canada Council and the support of an operating
grant from the Natural Sciences and Engineering Research Council of Canada.

\vfil\eject

\everypar= { \parindent=0pt \parskip=0pt \hangindent=24 pt \hangafter=1 }
\centerline{REFERENCES} 
\nobreak
\noindent

Bell,~R.~A., Briley,~M.~M., \& Norris,~J.~E. 1992, AJ, 104, 1127

Bencivenni,~D., Caputo,~F., Manteiga,~M., \& Quarta,~M.L. 1991,
ApJ, 380, 484

Bergbusch,~P.~A. 1993, AJ, 106, 1024

Bergbusch,~P.~A., \& VandenBerg,~D.~A. 1992, ApJS,
81, 163

Binney,~J. 1976, MNRAS, 177, 19

Bolte,~M. 1989, AJ, 97, 1688

Bolte,~M. 1992, ApJS, 82, 145

Bolte,~M., \& Dubath,~P. 1996, in preparation
Brown,~J.~A., \& Wallerstein,~G. 1993, AJ, 106, 142

Buonanno,~R. 1993, in {\it The Globular Cluster--Galaxy Connection\/},
eds. G.~H.~Smith and J.~P.~Brodie, ASP Conf. Ser., 48, 131

Buonanno,~R., Buscema,~G., Fusi~Pecci,~F., Richer,~H.~B., \&
Fahlman,~G.~G. 1990, AJ, 110, 1811

Buonanno,~R., Caloi,~V., Castellani,~V., Corsi,~C.~E., Fusi~Pecci,~F.,
and Gratton,~R. 1986, A\&A, 66, 79

Buonanno,~R., Corsi,~C.~E., \& Fusi~Pecci,~F. 1985, A\&A, 97, 117

Buonanno,~R., Corsi,~C.~E., \& Fusi~Pecci,~F. 1989, A\&A, 216, 80

Buonanno,~R., Corsi,~C.~E., Fusi~Pecci,~F., Fahlman,~G.~G., \&
Richer,~H.~B. 1994, ApJL, 430, L121

Buonanno,~R., Corsi,~C.~E., Fusi~Pecci,~F., Richer,~H.~B., \&
Fahlman,~G.~G. 1993, AJ, 105, 184

Buonanno,~R., Corsi,~C.~E., Pulone,~L., Fusi~Pecci,~F., Richer,~H.~B.,
\& Fahlman,~G.~G. 1995, AJ, 109, 663

Burkert,~A., Truran,~J.~W., \& Hensler,~G. 1992, ApJ, 391, 651

Caldwell,~S.~P., \& Dickens,~R.~J. 1988, MNRAS, 234, 87

Caloi,~V., Castellani,~V., \& Tornamb\`e,~A. 1978, A\&AS, 33, 169

Caputo,~F., \& Degl'Innocenti,~S. 1995, A\&A, 298, 833

Castellani,~V., \& Tornamb\`e,~A. 1977, A\&A, 61, 427

Catelan,~M., \& de~Freitas~Pacheco,~J.~A. 1993, AJ, 106, 1858

Catelan,~M., \& de~Freitas~Pacheco,~J.~A. 1994, A\&A, 289, 394

Catelan,~M., \& de~Freitas~Pacheco,~J.~A. 1995, A\&A, 297, 345

Chaboyer,~B., \& Demarque,~P. 1994, ApJ, 433, 510

Chaboyer,~B., Demarque,~P., \& Sarajedini,~A. 1996, ApJ, 459, 558\quad (CDS)

Chaboyer,~B., Sarajedini,~A., and Demarque,~P. 1992, ApJ, 394, 515

Crocker,~D.~A., \& Rood,~R.~T. 1984, in {\it Observational Tests of
the Stellar Evolution Theory\/}, IAU Symp.\ 105, eds. A.~Maeder and 
A.~Renzini (Dordrecht: Reidel), 159

Da~Costa,~G.~S., \& Armandroff,~T.~E. 1990, AJ, 100, 162

Da~Costa,~G.~S., Armandroff,~T.~E., \& Norris,~J.~E. 1992, AJ, 104, 154

Demarque,~P., \& Hirshfeld,~A. 1975, ApJ, 202, 346

Dickens,~R.~J., Croke,~B.~F.~W., Cannon,~R.~D., \& Bell,~R.~A. 1991,
Nature, 351, 212

Djorgovski,~S., \& Meylan,~G. 1994, AJ, 108, 1292

Durrell,~P.~R., \& Harris,~W.~E. 1993, AJ, 105, 1420

Eggen,~O.~J., Lynden-Bell,~D., \& Sandage,~A. 1962, ApJ, 136, 748\quad (ELS)

Faulkner,~J. 1966, ApJ, 144, 978

Ferraro,~I., Ferraro,~F.~R., Fusi~Pecci, F.,~Corsi,~C.~E., \&
Buonanno,~R. 1995, MNRAS, 275, 1057

Frogel,~J.~A., Persson,~S.~E., \& Cohen,~J.~G. 1981, ApJ, 246, 842

Frogel,~J.~A., Persson,~S.~E., \& Cohen,~J.~G. 1983, ApJS, 53, 713

Fullton,~L.~K., Carney,~B.~W., Olszewski,~E.~W., Zinn,~R., \& Demarque,~P.,
Janes,~K.~A., Da~Costa,~G.~S., \& Seitzer,~P. 1995, AJ, 110, 652

Fusi~Pecci,~F., Bellazzini,~M, Cacciari,~C., Ferraro,~F.~R. 1995, AJ,
110, 1664

Fusi~Pecci,~F., Ferraro,~F., Bellazzini,~M., Djorgovski,~S., Piotto,~G., 
\& Buonanno,~R. 1992, AJ, 105, 1145

Fusi~Pecci,~F., Ferraro,~F.R., Crocker,~D.~A., Rood,~R.~T., \&
Buonanno,~R. 1990, A\&A, 238, 95

Gratton,~R.~G. 1985, A\&A, 147, 169

Gratton,~R.~G., \& Ortolani,~S. 1988, A\&AS, 73, 137

Green,~E.~M., Demarque,~P., \& King,~C.~R. 1987, {\it The Revised Yale
Isochrones and Luminosity Functions\/}, (New Haven: Yale Univ. Obs.)

Green,~E.~M., \& Norris,~J.~E. 1990, ApJL, 353, L17

Grundahl,~F., 1996, Ph.D.\ dissertation, University of Aarhus

Harris,~W.~E. 1982, ApJS, 50, 573

Heasley,~J.~N., \& Christian,~C.~A.  1991, AJ, 101, 967

Hesser,~J.~E., Harris,~W.~E., VandenBerg,~D.~A., Allwright,~J.~W.~B.,
Shott,~P., and Stetson,~P.~B. 1987, PASP, 99, 739

Ibata,~R.~A., Gilmore,~G., \& Irwin,~M.~J. 1994, Nature, 370, 194

Kant,~I. 1755, {\it Universal Natural History and Theory of the
Heavens\/}, ed.\ and trans.\ W.~Hastie (1969; Ann Arbor: University of
Michigan Press)

Kaufman,~M. 1975, Ap.SpaceSci., 33, 265

Kinman,~T.~D. 1959, MNRAS, 119, 528

Kraft,~R.~P., Sneden,~C., Langer,~G.~E., \& Prosser,~C.~F. 1992, AJ, 104,
645

Kubiak,~M. 1991, {\it Acta Astronomica\/}, 41, 231

Lee,~S.~W. 1977, A\&AS, 27, 381

Lee,~Y.-W. 1991, in {\it The Formation and Evolution of Star
Clusters\/}, ed. K.~Janes,\break ASP Conf. Ser., 13, 205

Lee,~Y.-W. 1992, PASP, 104, 798

Lee,~Y.-W. 1993, in {\it The Globular Cluster--Galaxy Connection\/},
eds. G.~H.~Smith and J.~P.~Brodie, ASP Conf. Ser., 48, 142

Lee,~Y.-W., Demarque,~P., \& Zinn,~R. 1988, in {\it Calibration of
Stellar Ages\/},\break ed.\ A.~G.~D.~Philip (Schenectady: L. Davis Press), 149

Lee,~Y.-W., Demarque,~P., \& Zinn,~R. 1994, ApJ, 423, 248

Lin,~D.~N.~C., \& Richer,~H.~B. 1992, ApJL, 388, L57

Mayall,~N.~U., 1946, ApJ, 104, 290

Norris,~J. 1981, ApJ, 248, 177

Norris,~J. 1983, ApJ, 272, 245

Olszewski,~E.~W., Canterna,~R., and Harris,~W.~E. 1984, ApJ, 281, 158

Oort,~J.~H. 1958, Spec.Vat.Ric.Ast., 5, 415

Ortolani,~S. 1987, in {\it Proc.\ ESO Workshop on Stellar 
Evolution and Dynamics in the Outer Halo of the Galaxy\/},
eds. M.~Azzopardi and F.~Matteucci (Garching: ESO), 341

Ortolani,~S., Renzini,~A., Gilmozzi,~R., Marconi,~G., Barbuy,~B., Bica,~E. \&
Rich,~R.~M. 1995, Nature, 377, 701

Peebles,~P.~J.~E., and Dicke,~R.~H. 1968, ApJ, 154, 891

Peterson,~C.~J. 1987, PASP, 99, 1153

Pound,~M.~W., Janes,~K.~A., and Heasley,~J.~N. 1987, AJ, 94, 1185

Renzini,~A. 1977, in {\it Advanced Stages of Stellar Evolution\/},
eds. P.~Bouvier and A.~Maeder, (Geneva: Geneva Observatory), 151 

Renzini,~A. 1983, Mem.Soc.Astron.Ital., 54, 335

Richer,~H.~B., Harris,~W.~E., Fahlman,~G.~G., Bell,~R.~A., Bond,~H.~E.,
Hesser,~J.~E., Holland,~S., Pryor,~C., Stetson,~P.~B., VandenBerg,~D.~A.,
\& van~den~Bergh,~S. 1966, submitted to ApJ

Rood,~R.~T. 1973, ApJ, 184, 815

Rood,~R.~T., Crocker,~D.~A., Fusi~Pecci,~F., Ferraro,~F.~R.,
Clementini,~G., \& Buonanno,~R. 1993, in {\it The Globular 
Cluster--Galaxy Connection}, eds. G.~H.~Smith and J.~P.~Brodie, ASP Conf.
Ser., 48, 218

Rood,~R.~T., \& Iben,~I.,~Jr. 1968, ApJ, 154, 215

Sandage,~A. 1953, AJ, 58, 61

Sandage,~A. 1982, ApJ, 252, 553

Sandage,~A. 1990, JRASC, 84, 70

Sandage,~A., Katem,~B., \& Sandage,~M. 1981, ApJS, 46, 41

Sandage,~A., \& Wildey,~R. 1967, ApJ, 150, 469

Sandquist,~A., Bolte, M., \& Stetson, P. B. 1996, ApJ, submitted

Sarajedini,~A., \& Demarque,~P. 1990, ApJ, 365, 219\quad (SD)

Sarajedini,~A., \& King,~C.~R. 1989, AJ, 98, 1624\quad (SK)

Sarajedini,~A., \& Lederman,~A. 1991, in {\it The Formation and
Evolution of Star Clusters\/}, ed. K.~Janes, ASP Conf.
Ser., 13, 293

Searle,~L. 1975. in {\it The Evolution of Galaxies and Stellar
Populations\/}, eds. B.~M.~Tinsley and R.~B.~Larson (New Haven: Yale
Univ. Obs.), 219

Searle,~L., \& Zinn,~R. 1978, ApJ, 225, 357\quad (SZ)

Simoda,~M., \& Iben,~I.,~Jr. 1970, ApJS, 22, 81

Smecker-Hane,~T.~A., Stetson,~P.~B., Hesser,~J.~E., \& Lehnert,~M.~D.
1994, AJ, 108, 507

Sneden,~C., Kraft,~R.~P., Langer,~G.~E., \& Shetrone,~M.~D. 1994, AJ,
107, 1773

Sneden,~C., Kraft,~R.~P., Prosser,~C.~F., \& Langer,~G.~E. 1991, AJ, 102,
2001

Sneden,~C., Kraft,~R.~P., Prosser,~C.~F., \& Langer,~G.~E. 1992, AJ,
104, 2121

Stetson,~P.~B. 1989, PASP, 102, 932

Stetson,~P.~B. 1993, in {\it The Globular Cluster--Galaxy Connection},
eds. G.~H.~Smith and J.~P.~Brodie, ASP Conf. Ser., 48, 14

Stetson,~P.~B., Hesser,~J.~E., VandenBerg,~D.~A., Bolte,~M., \&
Smith,~G.~H. 1988, in {\/The Calibration of Stellar Ages\/}, ed.
A.~G.~D.~Philip (Schenectady: L. Davis Press), 19

Stetson,~P.~B., \& Smith,~G.~H. 1987, in {\it Proc.\ ESO Workshop on
Stellar Evolution and Dynamics in the Outer Halo of the Galaxy\/},
eds. M.~Azzopardi and F.~Matteucci (Garching: ESO), 387

Stetson,~P.~B., \& VandenBerg,~D.~A. 1996, in preparation.

Stetson,~P.~B., VandenBerg,~D.~A., Bolte,~M., Hesser,~J.~E., \&
Smith,~G.~H. 1989, AJ, 97, 1360

Sweigart,~A.~V., \& Gross,~P.~G. 1978, ApJS, 36, 405

VandenBerg,~D.~A. 1992, ApJ, 391, 685

VandenBerg,~D.~A., Bolte,~M., \& Stetson,~P.~B. 1990, AJ, 100, 445\quad (VBS)

VandenBerg,~D.~A., Bolte,~M., \& Stetson,~P.~B. 1996, to appear in ARAA

VandenBerg,~D.~A., \& Durrell,~P.~R. 1990, AJ, 99, 221

van~den~Bergh,~S. 1965, JRASC, 59, 151

van~den~Bergh,~S. 1967, AJ, 72, 70

van~den~Bergh, S.~1993, AJ, 105, 971

von~Weizsacker,~C.~F. 1955, ZfAp, 35, 252

Walker,~A.~R. 1992, PASP, 94, 1063

Zinn,~R. 1985, ApJ, 293, 424

Zinn,~R. 1993, in {\it The Globular Cluster--Galaxy Connection},
eds. G.~H.~Smith and J.~P.~Brodie, ASP Conf. Ser., 48, 38

\vfil\eject

\centerline{FIGURE CAPTIONS} \medskip \nobreak 
\noindent Fig.~1.\quad Bergbusch \& VandenBerg (1992) isochrones for [Fe/H]
$=-2.03$ and $-1.26$ and the ages indicated are plotted as solid curves
from the bottom center to the upper right part of the figure, and the main
sequence (MS), subgiant branch (SGB), and red-giant branch (RGB)
evolutionary stages are labeled along the more metal-rich loci.  
For each metal abundance, the youngest age has the brightest, bluest
turnoff and the bluest giant branch.  Fully consistent zero-age horizontal
branches (ZAHB) as given by Dorman (1992) are plotted as dashed curves,
with the longer curve representing the ZAHB for the [Fe/H] = --2.03 case
while, for the sake of clarity, only the red end of Dorman's ZAHB for
[Fe/H] $=-1.26$ is shown (the lower of the two dashed loci).  Solid curves
rising from the metal-poor ZAHB locus represent a selection of
horizontal-branch evolutionary tracks for different post-giant-branch
stellar masses.  The approximate location of the RR Lyrae instability strip
is indicated by the parallel dotted lines.  The meaning of the quantity
\delv\ --- which represents the magnitude difference between the bluest
point at the turnoff (TO) and the horizontal branch (HB) at the same color
--- is illustrated for the case [Fe/H] = --2.03.  This quantity is
clearly a function of age and somewhat insensitive to [Fe/H] given that the
luminosity of the horizontal branch and of the turnoff point (at a fixed
age) are both predicted to decrease as the metallicity increases (though
not at exactly the same rate).

Fig.~2.\quad Panel (a): VandenBerg \etal\ (1996) isochrones for the
specified parameters and ages in 2~Gyr increments from 12 to 18~Gyr (in
the order indicated), after they have been registered so that they have
the same turnoff colors and identical magnitudes at the point on their
respective lower main sequences that is 0.05 mag redder than the turnoff.
Panel (b): Registration of the \M{68}, \M{30}, and \ngc{6397} fiducials to
that of \M{92} and their superposition onto the lower RGB segments of the
isochrones plotted in panel (a).  The sources of the data and the
horizontal and vertical offsets that were applied to the cluster
photometry are given by VandenBerg \etal\ (1990).

Fig.~3.\quad As for Fig.~2; in this case, the isochrones assume a value of
$\alpha_{\rm MLT} = 2.5$, but are otherwise identical to those portrayed
in the previous figure.

Fig.~4.\quad A demonstration of how the shape of a 16 Gyr isochrone for
[Fe/H] $=-2.14$, [$\alpha$/Fe] $=0.3$, $Y=0.235$, and $\alpha_{\rm MLT} =
1.89$ would change, according to VandenBerg \etal\ (1996) models, if the
latter four parameters were varied by modest amounts (as indicated).  The
dashed and dot-dashed curves were registered to the solid curves at the
magnitudes and colors indicated by the small arrows.

Fig.~5.\quad HB morphology vs. [Fe/H] for globular clusters having $R_G <
8$~kpc (closed circles), $8\le R_G\le 40$~kpc (open circles), and $R_G >
40$~kpc (crosses).  The data and the solid lines, which represent
theoretical isochrones produced by a synthetic HB code, are due to Lee
\etal\ (1994).  The clusters that have been identified on this plot are
discussed in the text.

Fig.~6.\quad Comparison of the turnoff regions of the observed fiducial
sequences for \ngc{288} (Bolte 1992), \ngc{362} (VandenBerg et al. 1990),
and \ngc{1851} (Walker 1992b), after vertical offsets of $+0.13$ and
$-0.60$ mag and color shifts of $-0.018$ and $-0.035$ mag, respectively,
were added to the tabulated sequences of the latter two clusters.  Due to
the obvious erratic difference among the sequence shapes, the vertical
registration was effected at the middle of the nearly horizontal part of
the subgiant branch.  As this part of the cluster locus is highly
sensitive to age, any age differential among the clusters should cause a
disregistration of their HBs when these offsets are applied.

Fig.~7.\quad Walker's (1992b) photometry for the red-giant and
horizontal-branch stars in \ngc{1851}.  Hand-drawn envelopes to the
stellar distributions are delineated by the solid curves.

Fig.~8.\quad Overlay of the solid curves from the previous figure onto the
Harris (1982) photometry of the bright stars in \ngc{362} and that by
Bergbusch (1993) for \ngc{288}, assuming exactly the same offsets as those
adopted in producing Fig.~6.  Zero-point shifts of $-0.010$ mag in $B-V$
and $+0.0635$ mag in $V$ were applied to the Bergbusch observations, given
that he has found that these corrections are required to place his
observations onto the system of the Bolte (1992) data, which were used to
effect the registration shown in Fig.~6.  At the red end of \ngc{288}'s
blue horizontal branch, stars appear to hug the upper envelope of the
\ngc{1851} horizontal branch; this may be because these stars have evolved
from bluer zero-age horizontal-branch positions than the ones in
\ngc{1851}, along tracks like those shown in Fig.~1.  
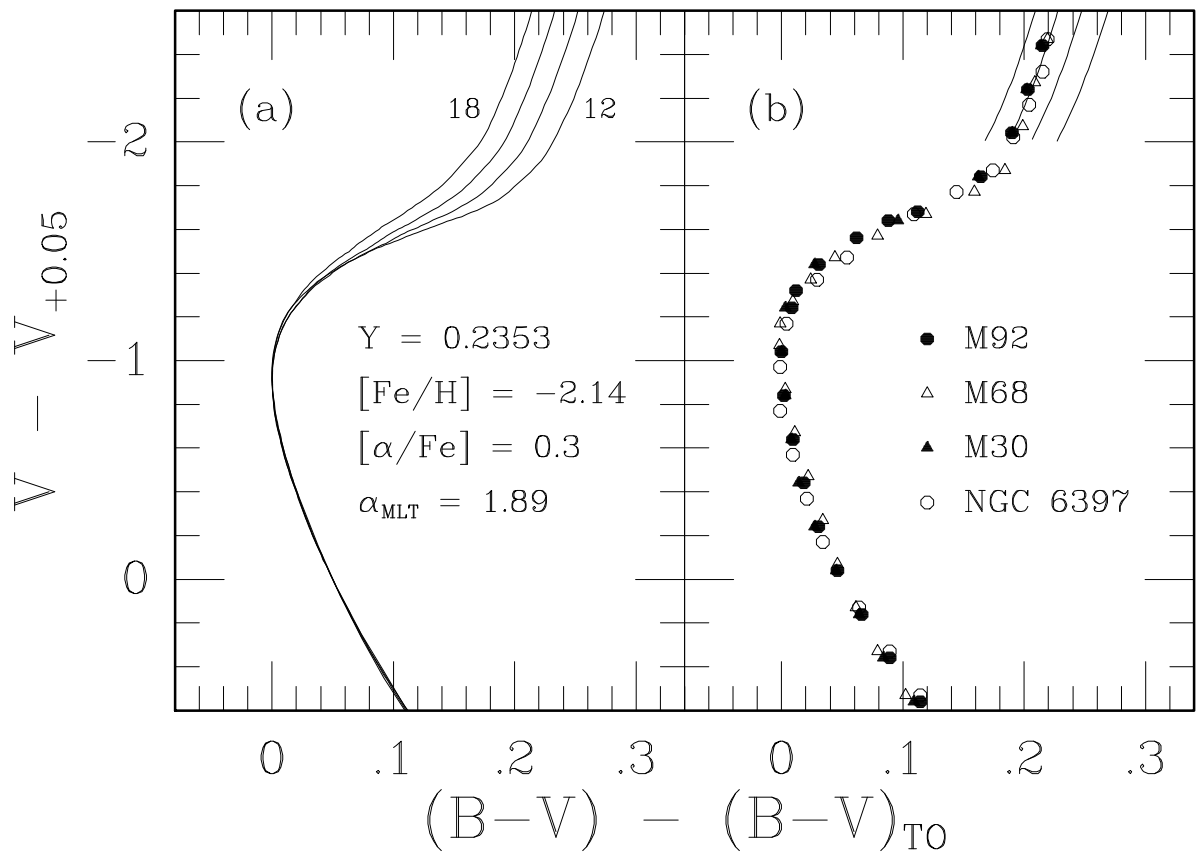
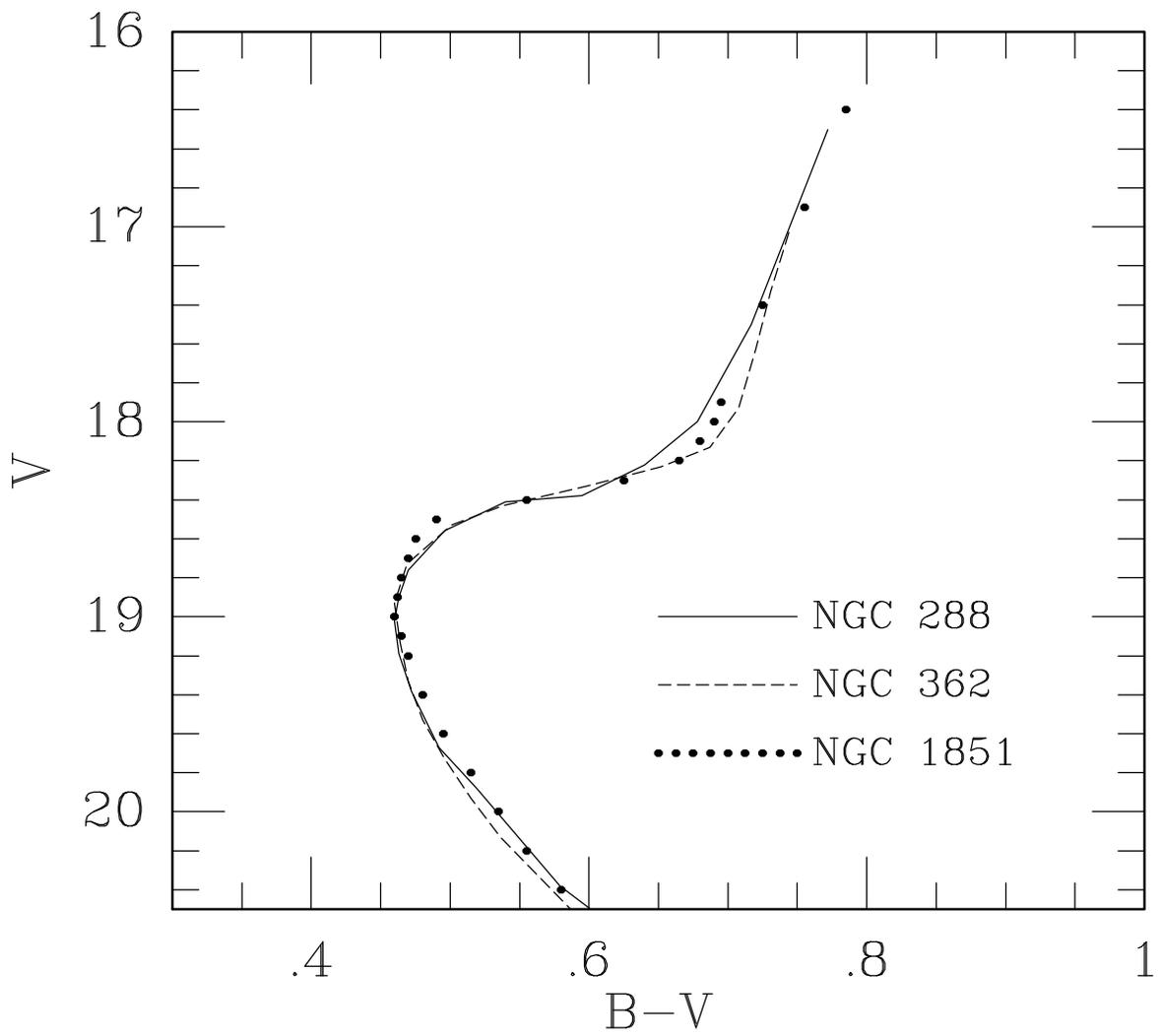
\bye